\documentclass[pra
	       ,nofootinbib
	       ,floatfix
	       ,superscriptaddress
	       ,twocolumn
	       ]{revtex4-1}
	       
\usepackage{times}

\usepackage{soul}
\usepackage{amsmath} 
\usepackage{amssymb}
\usepackage{cool}
\usepackage{graphicx} 
\usepackage{color} 
\usepackage[utf8]{inputenc}
\usepackage[hidelinks]{hyperref}
\usepackage[hidelinks]{hyperref}
\usepackage{multirow}
\usepackage{siunitx}

\newcommand{\SOMfirstPICposition}{\begin{figure}[b]}

\begin{document}

\title{Relaxation to a Phase-locked Equilibrium State in a One-dimensional Bosonic Josephson Junction}

\author{Marine~Pigneur}
\affiliation{Vienna Center for Quantum Science and Technology, Atominstitut, TU Wien, Stadionallee 2, 1020 Vienna, Austria}

\author{Tarik~Berrada} 
\author{Marie~Bonneau}   
\author{Thorsten~Schumm}   
\affiliation{Vienna Center for Quantum Science and Technology, Atominstitut, TU Wien, Stadionallee 2, 1020 Vienna, Austria}
\author{Eugene Demler}
\affiliation{Department of Physics, Harvard University, Cambridge, Massachusetts 02138, USA}
\author{J\"org~Schmiedmayer}
\email[]{schmiedmayer@atomchip.org}
\affiliation{Vienna Center for Quantum Science and Technology, Atominstitut, TU Wien, Stadionallee 2, 1020 Vienna, Austria}

\date{\today}

\begin{abstract}
We present an experimental study on the non-equilibrium tunnel dynamics of two coupled one-dimensional Bose-Einstein quasi-condensates deep in the Josephson regime. Josephson oscillations are initiated by splitting a single one-dimensional condensate and imprinting a relative phase between the superfluids. Regardless of the initial state and experimental parameters, the dynamics of the relative phase and atom number imbalance shows a relaxation to a phase-locked steady state. The latter is characterized by a high phase coherence and reduced fluctuations with respect to the initial state. We propose an empirical model based on the analogy with the anharmonic oscillator to describe the effect of various experimental parameters. A microscopic theory compatible with our observations is still missing.
\end{abstract}

\maketitle

All quantum evolution is in principle unitary, and thus isolated quantum systems should never relax to a steady state. However, both theoretical and experimental works \cite{Langen2015r,Polkovnikov2011,Calabrese2015,Eisert2015} show that non-integrable systems reach a relaxed state resembling a Gibbs ensemble \cite{Srednicki1994,Rigol2008}.  
In contrast, integrable systems generally relax to pre-thermal steady states, for which a description by a generalized Gibbs ensemble reflects the conserved quantities in the system. It is especially interesting to study the thermalization in systems that are close to being integrable. 

The recent experimental advances in manipulating and probing ultracold atomic gases established them as ideal model system to study the non-equilibrium dynamics and relaxation of isolated quantum systems \cite{Langen2015r}. In this context, an interesting model system is a bosonic Josephson junction consisting of two coupled superfluids \cite{Levy2007, Gati2007}. Its dynamics in the three-dimensional case has been investigated experimentally \cite{Albiez2004,Leblanc2011,Spagnolli2017}. The physics in reduced dimensions (1D) can be essentially described by the quantum Sine-Gordon model \cite{Gritsev07, Thirring195891,Coleman75,Mandelstam,Faddeev19781,Sklyanin1979}, as recently verified for systems in thermal equilibrium \cite{Schweigler2017}. The integrability of this model implies rich dynamics. In particular, the dynamics occurring after a quench of the tunnel coupling was recently studied in a series of theoretical papers \cite{Torre2013,Foini2015,Foini2017} and showed a slow and incomplete relaxation. In this letter, we present an experimental study of the dynamics of two tunnel-coupled 1D-superfluids initialized with equal numbers and a uniform relative phase $0 < \phi_0 < \pi$.  The system relaxes to a phase-locked equilibrium state. 

\begin{figure}[ht!] 
\includegraphics[width=\columnwidth]{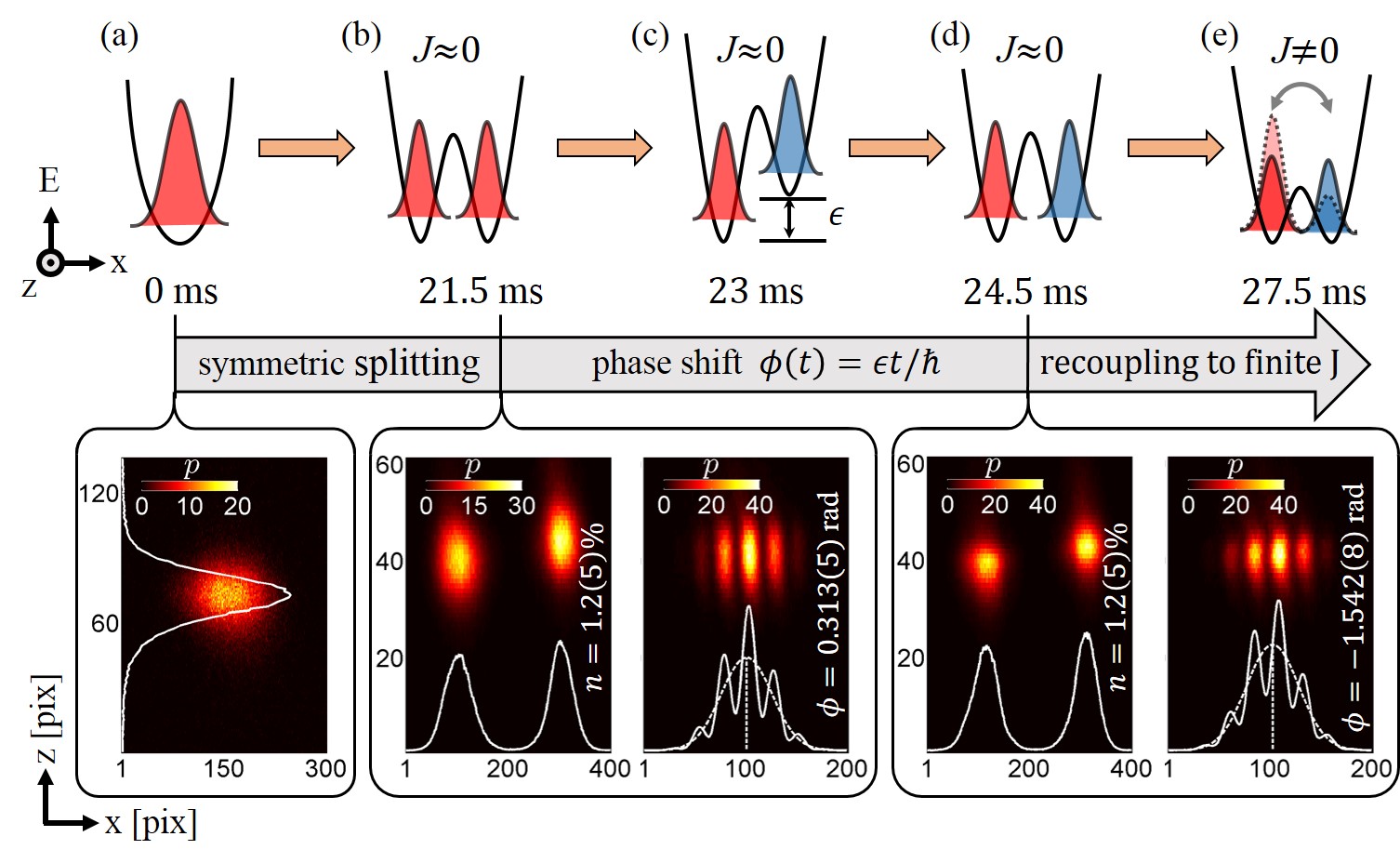}
\caption{\label{Fig:sequence} Schematic of the preparation sequence consisting in a splitting \textbf{(a,b)} of the original wave-packet to a trap of negligible coupling, a phase imprinting between the wave-packets \textbf{(c,d)} and a recoupling to initiate a tunneling dynamics \textbf{(e)}. The captions display the fluorescence pictures (averaged over 10 repetitions) of the atomic density after \SI{46}{ms} time-of-flight and the corresponding integrated profiles.} 
\end{figure} 

Our experimental system consists of two one-dimensional quasi-condensates (1D-BEC) of ${^{87}}$Rb magnetically trapped in a double-well potential with tunable barrier height. The preparation protocol, illustrated by Fig.~\ref{Fig:sequence}, relies on an atom chip \citep{Trinker2008} to coherently manipulate the wave-packets \cite{Berrada2013}. The sequence can be interrupted at any stage to perform fluorescence imaging \cite{Bucker2009} after a time-of-flight of \SI{46}{ms}. Fluorescence pictures of the main stages are displayed in Fig.~\ref{Fig:sequence}. 
We start the sequence with a single 1D-BEC obtained by evaporative cooling in an trap elongated along the $z$-axis (Fig.~\ref{Fig:sequence}(a)). The initial condensate contains $750$ to $4500$\,atoms. The trap frequencies are ${\omega_{x,y}=2\pi\times}$\SI{3}{kHz} transversely and ${\omega_z=2\pi\times}$\SI{22}{Hz} longitudinally, corresponding to a condensate length between \SI{20}{\micro m} (for $750$\,atoms) and \SI{33}{\micro m} (for ${4500}$\,atoms) \cite{Gerbier2004}. Yang-Yang thermometry \cite{Davis2012} gives an estimate of the initial temperature of $T=\SI{18(3)}{nK}$. 

Using radiofrequency-dressing \cite{Lesanovsky2006,Hofferberth2006}, we deform the trap in \SI{21.5}{ms} by continuously raising a barrier and obtain a double-well potential elongated along the $z$-axis and symmetric with respect to the barrier (Fig.~\ref{Fig:sequence}(b)). Along the $x$-axis, it results in a splitting of the wave-function during the last \SI{5}{ms} of the ramp. The barrier is made high enough to neglect tunneling, such that the two 1D-BECs are considered decoupled. We define the relative phase
\begin{equation}
\phi(t)=\phi_L(t)-\phi_R(t),
\end{equation} 
with $\phi_{L,R}(t)$ the phase of the left and right component, respectively. The phase is experimentally extracted from the interference pattern resulting from the overlap of the wave-functions after time-of-flight (cf. pictures of Fig.\ref{Fig:sequence}(b,d)). The conjugated variable is the normalized atom number imbalance defined by:
\begin{equation}
n(t)=\frac{N_L(t)-N_R(t)}{N_L(t)+N_R(t)},
\end{equation}
with $N_{L,R}(t)$ the atom number of the left and right component, respectively. The imbalance measurement requires to move the clouds further apart by raising the barrier height. This prevents their overlap in time-of-flight. At this stage, the averaged values of both the imbalance and the phase are close to zero (cf. pictures of Fig.~\ref{Fig:sequence}(b,d)) due to the trap symmetry. 

We then imprint a initial relative phase $\phi_0$ by shifting one site of the double-well along the vertical $y$-axis in \SI{1.5}{ms} (Fig.~\ref{Fig:sequence}(c)). This introduces an energy difference $\epsilon$ between the two sites and results in a phase accumulation ${\phi(t)=\epsilon t/\hbar}$. The trap symmetry is then re-established in \SI{1.5}{ms} (Fig.~\ref{Fig:sequence}(d)). We prepare a relative phase between $0$ and $\pi$ by varying the value of the vertical shift. The phase appears as a shift between the integrated profile maximum and the center of the envelope, as displayed in the fluorescence pictures of Fig.~\ref{Fig:sequence}(d) and Fig.~\ref{Fig:fringes}(a)). The straightness of the fringes shows that the relative phase is uniformly imprinted along the elongated direction of the condensate with negligible fluctuations at the scale of the imaging resolution (\SI{4}{\micro m} in object space) \cite{Bucker2009}. In the decoupled trap, the relative phase randomizes under the effect of interaction-induced phase diffusion \cite{Lewenstein1996,Javanainen1997,Leggett1998}. In our case, this effect is strongly reduced by a large number-squeezing factor obtained by the splitting of the initial BEC \cite{Berrada2013,Jo2007}. We define the number-squeezing factor by ${\xi_N=\Delta n/\sqrt{N}}$ with $\Delta n$ the standard deviation of the imbalance distribution and $N$ the total atom number. For our typical atom number $N=2500(200)$\,atoms, we obtain ${\xi_N=0.57(6)}$. The corresponding phase diffusion rate is \SI{0.05(2)}{rad/ms}. In circular statistics, the phase coherence is indicated by the phasor $R$ of the phase distribution, which varies between $0$ for a random distribution and $1$ for a perfectly narrow one. The phasor degrades from $R_\mathrm{split}= 0.94(2)$ after splitting to ${R_0=0.91(2)}$ at the end of the preparation sequence, indicating that a high phase coherence is preserved (Fig.~\ref{Fig:fringes}(a)) and that the initial phase $\phi_0$ is well defined. The imbalance is not affected by the phase shift such that its value is $n_0\approx 0$.

\begin{figure}[h] 
\includegraphics[width=\columnwidth]{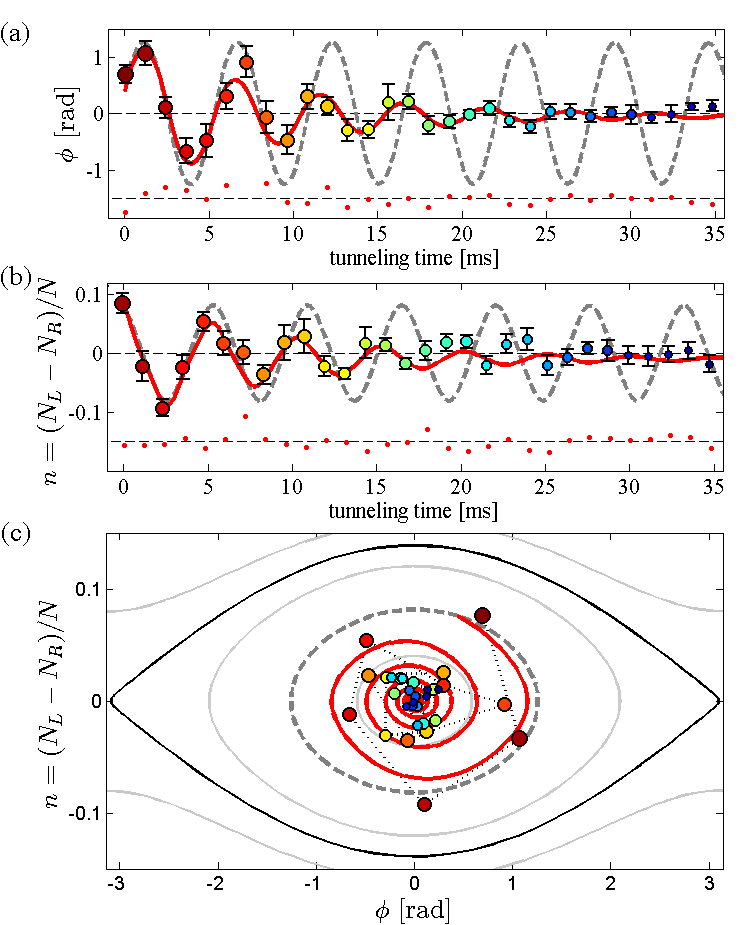}
\caption{\label{Fig:qualitative} Damped Josephson oscillations of the relative phase \textbf{(a)} and imbalance \textbf{(b)} for ${N=3300(600)}$\,atoms, ${\phi_0=\SI{-1.3(4)}{rad}}$ and ${n_0=-0.004(13)}$. Red line: Fit result giving a damping time ${\tau=\SI{9.8(2)}{ms}}$, ${U/h=\SI{0.71(15)}{Hz}}$ and ${J/h=\SI{8(2)}{Hz}}$. Red dots: residuals (shifted for clarity). Dashed line: corresponding predictions of the mean-field two-mode Bose-Hubbard model. \textbf{(c)} Evolution in the phase portrait representation.}
\end{figure}

Finally, the barrier is lowered in \SI {3}{ms} to reconnect the wave-packets (Fig.~\ref{Fig:sequence}(e)). This initiates the tunneling dynamics with a single particle tunnel-coupling strength varying between ${J/h=\SI{2(1)}{Hz}}$ and ${J/h=\SI{32(3)}{Hz}}$. The corresponding transverse frequencies vary between ${\omega_x=2\pi\times}$\SI{1.2}{kHz} and ${\omega_x=2\pi\times}$\SI{1.5}{kHz}. The smallest barrier height ${h_B/h=\SI{3.2}{kHz}}$ is higher than the largest chemical potential ${\mu/h=\SI{1.9}{kHz}}$ obtained for ${4500}$\,atoms \cite{Gerbier2004}. It shows that the observed dynamics primarily results from tunneling. The relative phase and imbalance evolve during a given tunneling time after which one of them is measured destructively. The preparation is repeated for increasingly long tunneling times to reconstruct the entire dynamics, alternating between phase and imbalance measurement. A typical example of the oscillating dynamics is displayed in  Fig.~\ref{Fig:qualitative} for ${N=3300(600)}$\,atoms, ${\phi_0=\SI{-1.3(4)}{rad}}$ and ${n_0=-0.004(13)}$. The oscillations of the mean phase and imbalance present a damping on a timescale of \SI{15}{ms} toward an equilibrium state (${n_{eq}\approx 0,\phi_{eq}\approx 0}$) without decrease of the total atom number. While the oscillations are expected from a two-mode Bose-Hubbard model \cite{Raghavan1999}, such a damping goes beyond the existing microscopic descriptions \cite{Torre2013,Foini2015,Sakmann2009}.

The analysis of the individual fluorescence images shows that the fringe patterns after damping are straight and centered on the envelope maximum, as displayed in Fig.~\ref{Fig:fringes}(b). We deduce that the relative phase reaches the value of zero uniformly along the condensate. Furthermore, the contrast of the integrated fringes barely degrades from ${C_0=0.56(7)}$ to ${C_{\text{eq}}=0.49(8)}$. This indicates that the longitudinal phase fluctuations do not increase at the timescale of the damping.

\begin{figure}[h] 
\includegraphics[width=\columnwidth]{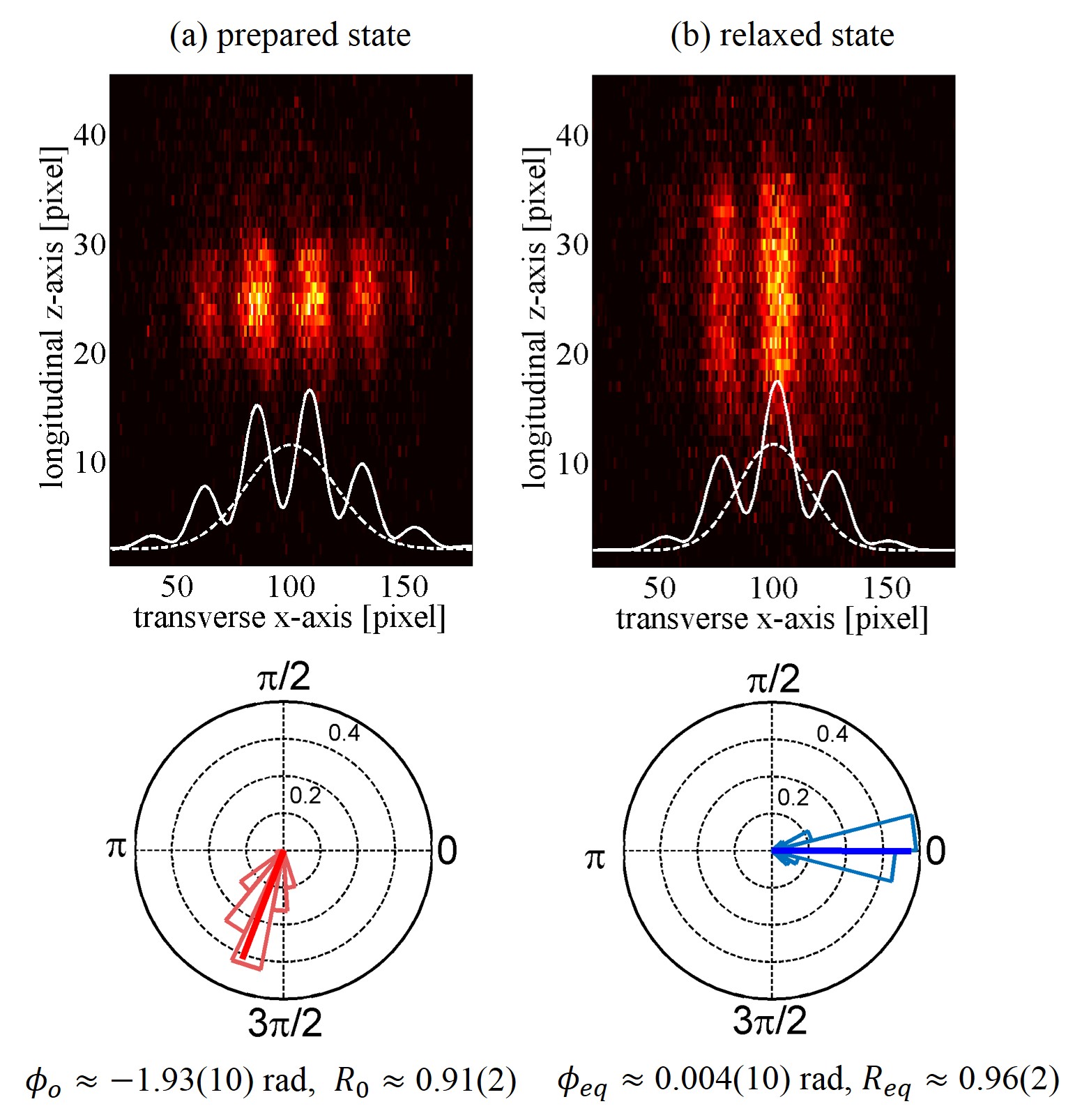}
\caption{\label{Fig:fringes} \textbf{(a) Top:} Single fluorescence picture of interference  fringes of the initial state and corresponding integrated profile. \textbf{Bottom:} Phase distribution (50 realizations) showing a high phase coherence. \textbf{(b) Top:} Single fluorescence picture of interference fringes obtained after relaxation and showing a uniform phase-locking. We observe an expansion of the cloud due to a slow breathing mode (see supplementary materials). \textbf{Bottom:} Phase distribution (50 realizations) showing an increase of phase coherence.}
\end{figure}

The phase distribution, obtained from 50 repetitions of an experimental sequence, shows that the phase locking is very reproducible (Fig.~\ref{Fig:fringes}(b)).  After relaxation, the phasor value reaches ${R_{\mathrm{eq}}=0.96(2)}$, such that the relaxation increases the phase coherence. The corresponding coherence factor is ${\left\langle\cos(\phi_\mathrm{eq})\right\rangle}=0.95(2)$. The phase-locking toward an equilibrium state dominates over dephasing phenomena expected in such a system \cite{Bouchoule2005}.

The relaxation is observed for a broad range of experimental parameters. In order to analyze the damped oscillations obtained under various conditions, we develop a phenomenological model adapted from the mean-field two-mode Bose-Hubbard model \cite{Gati2007}. We call $U$ the on-site interaction energy and $J$ the single particle tunnel coupling energy. The ratio $NU/2J$ is in the order of 100, placing our experiment deeply into the Josephson regime \cite{Leggett2001}. The  derivation of the Hamiltonian for a symmetric trap gives the following undamped time evolution of the phase and imbalance:
\begin{align}
\dot{n}(t)&\approx-\frac{2J}{\hbar}\sqrt{1-n^2(t)}\sin\left(\phi(t)\right),\label{eq:n_dot}\\
\dot{\phi}(t)&\approx\frac{NU}{\hbar} n(t)+\frac{2J}{\hbar}\frac{n(t)}{\sqrt{1-n^2(t)}}\cos\left(\phi(t)\right). \label{eq:phi_dot}
\end{align}

The interplay between the inter-atomic interaction $NU$ and the tunneling $2J$ leads to different dynamical modes. As demonstrated in \cite{Raghavan1999}, every initial state (${n(0),\phi(0)}$) obeying  
\begin{equation}
\frac{NU}{4J}n(0)^2-\sqrt{1-n(0)^2}\cos(\phi(0))\leq1
\label{eq:separatrix}
\end{equation}
results in Josephson oscillations of the phase and imbalance characterized by the  plasma frequency $\omega_p\approx\sqrt{NU2J}/\hbar$. Our experimental protocol prepares ${n(0)\approx 0}$ such that Eq.~(\ref{eq:separatrix}) is always verified. The system presents an analogy with a classical momentum-shortened pendulum \cite{Marino1999} in which the relative phase $\phi$ is analogous to the pendulum angle and the imbalance $n$ is proportional to its momentum $\dot{\phi}$. In \cite{Marino1999}, the length of the pendulum is defined by ${l(t)=\sqrt{1-n^2(t)}}$. Defining $N_0$ as the amplitude of the $n-$oscillations, it follows from Eq.~(\ref{eq:separatrix}) that ${N_0\leq2\sqrt{2J/NU}}$ in the limit of $NU\gg2J$. Consequently, the momentum-shortening is negligible in our case and the analytical solution of a rigid pendulum expressed in terms of the sn-Jacobi elliptic function is a good approximation \cite{Ochs2011,Abramowitz1972}. To account for the damping of the oscillations, we follow the approach of \cite{Marino1999} and add a dissipative term to Eq.~(\ref{eq:n_dot}):
\begin{equation}
\dot{n}(t)\approx-\frac{2J}{\hbar}\sqrt{1-n^2(t)}\sin\left(\phi(t)\right) -\frac{\eta}{N}\dot{\phi}(t),\label{eq:n_damped_dot}
\end{equation}
where $\eta$ is an empirical dimensionless viscosity. $\eta$ normalized by $N$ has the physical meaning of the shunting conductance in the Resistively and Capacitively Shunted Junction (RCSJ) model, in which a damping appears as in Eq.~(\ref{eq:n_damped_dot}) \cite{Tinkham,Schon90}. The viscosity results in an exponential decay with the characteristic time $\tau$. In the harmonic regime and in the limit $NU\gg2J$, $\tau$ reads:
\begin{equation}
\tau(U)\approx\frac{2\hbar}{U\eta}. \label{eq:tau}
\end{equation}
Assuming that Eq.~(\ref{eq:tau}) holds true for large amplitude oscillations, we empirically modify the analytical solutions to account for a damping. The phase $\phi(t)$ and the imbalance $n(t)$ become:
\begin{eqnarray}\nonumber
\phi(t)&\approx& 2\arcsin \left[ \sin \left( \frac{\Phi_0}{2}\right)e^{-t/\tau}\right. \\ 
       &&   \left.\times\JacobiSN{\omega t+\phi'}{\sin\left(\frac{\Phi_0}{2}\right) e^{-t/\tau} } \right], \label{eq:phi_damped}\\
n(t)&\approx&\frac{N_0}{\omega\times 2\sin(\frac{\Phi_0}{2})}\dot{\phi}(t). \label{eq:n_damped}
\end{eqnarray}
$\Phi_0$ and $N_0$ are the amplitudes of the phase and  imbalance oscillations. 
$\omega$ is the damped harmonic frequency, which differs from the plasma frequency $\omega_p$ by ${\omega=\sqrt{\omega_p^2-1/\tau^2}}$. However, the correction introduced by $\tau$ is negligible. ${\phi'}$ is a temporal shift of the oscillations to compensate that the sn-function conventionally starts at $\phi(0)=\Phi_0$. The argument ${\sin\left(\frac{\Phi_0}{2}\right)e^{-t/\tau}}$ in the $\text{sn}$-function describes the anharmonicity of the oscillation, initially set by $\Phi_0$ and exponentially decreasing over time under the effect of $\eta$. 
\begin{figure*}[t] 
\includegraphics[width=\linewidth]{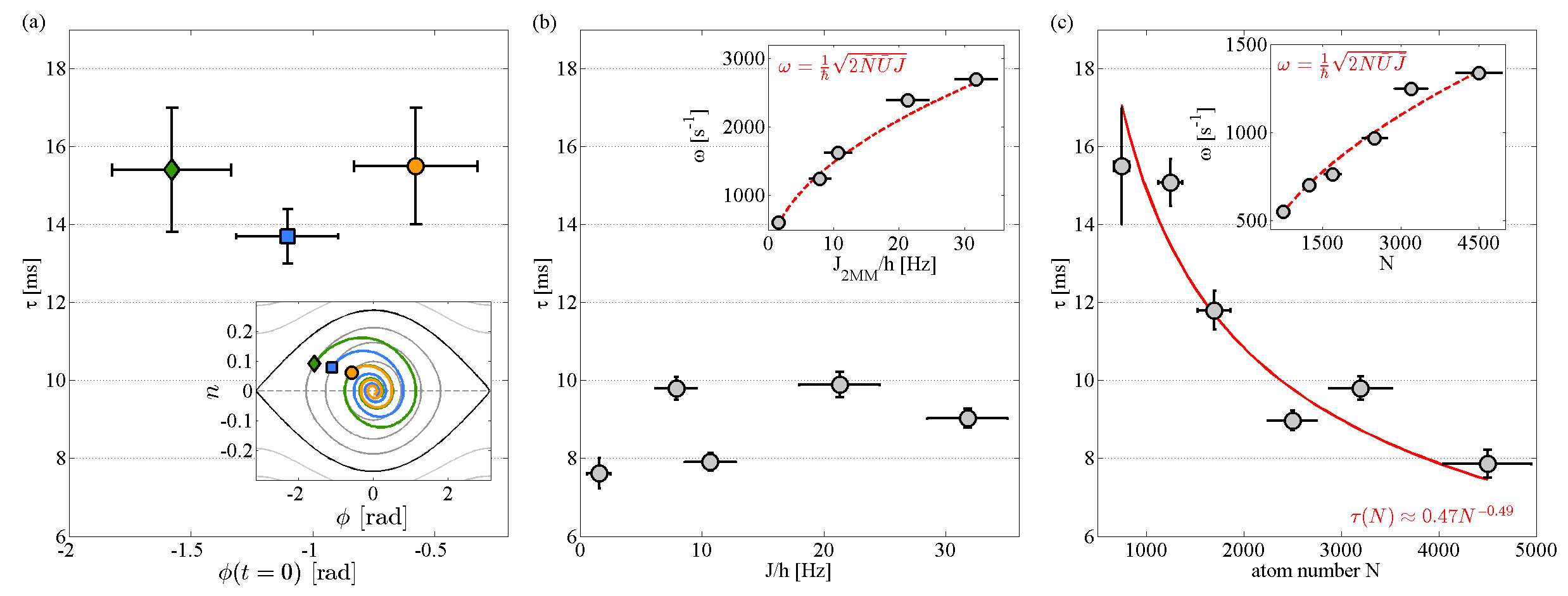}
\caption{\label{Fig:Parameters} \textbf{Relaxation dependence on parameters} \textbf{(a)} Variation of $\tau$ with the initial phase for $N=750(100)$\,atoms and $J/h=\SI{6(1)}{Hz}$. The damping time remains constant within the error bars, excluding a dependence on the oscillation amplitudes. Inset: Fit results for the various prepared phases showing the self-consistency of the trajectories. \textbf{(b)} Variation of $\tau$ with the tunnel coupling for $N=3500(500)$\,atoms and $\phi_0=\SI{1.93(35)}{rad}$. $\tau$ does not present a dependence on the tunnel coupling $J/h$ between \SI{2(1)}{Hz} and \SI{32(3)}{Hz}. Inset: Fit results for the plasma frequency $\omega$ and prediction of the 2-mode model for constant values of $\bar{N}=3500$\,atoms and $\bar{U}/h=\SI{0.85}{Hz}$ (red dashed line). \textbf{(c)} Variation of $\tau$ with the atom number $N$ for $J/h=\SI{7(2)}{Hz}$ and $\phi_0=\SI{-1.48(98)}{rad}$. $\tau$ presents a dependence in ${1/\sqrt{N}}$ (red line). Inset: $\omega$ versus N and prediction of the 2-mode model (dashed red line) for the averaged values $\bar{U}/h=\SI{0.71}{Hz}$ and $\bar{J}/h=\SI{7}{Hz}$ deduced from the fit parameters.}
\end{figure*}  
We establish a connection between $U, J$ and $\eta$ of the two-mode model and $N_0$,$\Phi_0$, $\omega$ and $\tau$ of the pendulum analogy. It relies on the approximation $\omega\approx\omega_p$, on the definition of $\tau$ given by Eq.~(\ref{eq:tau}) and by comparing Eq.~(\ref{eq:n_damped}) with the linearized Eq.~(\ref{eq:phi_dot}).  
\begin{eqnarray}
J&\approx&\frac{\hbar\omega}{4}\frac{N_0}{\sin(\Phi_0/2)},\label{eq:J}\\
U&\approx&2\hbar\omega\frac{\sin(\Phi_0/2)}{N\times N_0},\label{eq:U}\\
\eta&\approx&\frac{N\times N_0}{\tau\omega\sin(\Phi_0/2)}. \label{eq:eta}
\end{eqnarray}

We use Eqs.~(\ref{eq:phi_damped}),(\ref{eq:n_damped}) as a fit model. The small amplitude of the $n$-oscillations makes it difficult to estimate $N_0$. As $\eta$ depends on $N_0$ through Eq.~(\ref{eq:eta}), fitting the dynamics with Eqs.~(\ref{eq:phi_dot}),(\ref{eq:n_damped_dot}) is unreliable. In contrast, $\tau$ shows no correlations with $N_0$ and indicates clearly the effect of the experimental parameters on the relaxation mechanism. We recover the values of $U$ and $J$ using Eqs.~(\ref{eq:J}),(\ref{eq:U}) and insert them in Eqs.~(\ref{eq:n_dot}),(\ref{eq:phi_dot}) to build the phase portrait presented in Fig.~\ref{Fig:qualitative}(c) and in the inset of Fig.~\ref{Fig:Parameters}(a).

Eqs.~(\ref{eq:phi_damped}),(\ref{eq:n_damped}) reproduce the decay of the oscillations amplitude. Also, as the system gets closer to the harmonic regime, the oscillation frequency increases. It shows in Fig.~\ref{Fig:qualitative}(a),(b) when comparing the data and fit (red line) with the undamped prediction (dashed gray line) of the 2-mode model. The absence of structure in the residuals shows that it is justified to fit the dynamics with a unique damping time, similarly to a pendulum evolving in a medium of fixed viscosity. This implies that the relaxation does not depend on the amplitude of the oscillation. 

A more systematic check of the dependence of $\tau$ on the oscillation amplitude is performed by preparing oscillations of initial phase $\phi_0$ varied between $-0.2\pi$ and $-0.8\pi$. $J$ and $N$ are kept constant with ${J/h=\SI{6(1)}{Hz}}$ (fit result) and ${N=750(100)}$\,atoms. Fig.~\ref{Fig:Parameters}(a) displays the values of $\tau$ for the different initial phases and the inset shows the trajectories deduced from the fit in the phase portrait representation. We observe a constant damping time ${\tau=\SI{15(1)}{ms}}$, corresponding to ${\eta=26(7)}$. Since a larger initial phase also implies a larger amplitude of the imbalance oscillation, we can state that the relaxation is independent of the $n-$oscillation amplitude.

We also check the influence of the tunnel coupling strength between ${J/h=\SI{2(1)}{Hz}}$ and ${J/h=\SI{32(3)}{Hz}}$ (fit result). The initial state is characterized by ${N=3500(500)}$\,atoms, ${n_0=-0.004(13)}$, ${\phi_0=\SI{-1.93(35)}{rad}}$, ${C_0=0.53(7)}$ and  ${R_0=0.94(3)}$. The fit results for the plasma frequency $\omega$ are displayed in the inset of Fig.~\ref{Fig:Parameters}(b) and follows the prediction of the 2-mode model. Fig.~\ref{Fig:Parameters}(b) shows that the damping does not present an obvious dependence on $J$. This implies that the damping does not result from the tunneling dynamics of the wave-functions, nor from excitations to higher transverse modes. Additionally, the absence of dependence of $\tau$ on $\Phi_0$, $N_0$ and $J$ shows that the relaxation does not depend on the plasma frequency, that we can write as ${\omega=\frac{4J}{\hbar N_0}\sin(\frac{\Phi_0}{2})}$. This excludes a friction between the atoms and a thermal background gas as a cause for the damping. 

In the following, we measure the dynamics for a total atom number varied between $750$ and $4500$\,atoms to investigate the effect on the relaxation. The barrier height is kept identical and the change of atom number has a negligible impact on the tunnel coupling which remains ${J/h=\SI{7(2)}{Hz}}$. The prepared phase is $\phi_0=\SI{-1.48(98)}{rad}$, associated to a contrast ${C_0=0.55(7)}$ and a phasor ${R_0=0.81(3)}$. The typical initial imbalance for the largest atom number is ${n_0=0.006(47)}$. The variation of $\tau$ with the atom number is displayed in Fig.~\ref{Fig:Parameters}(c). We fit $\tau$ by the function $\alpha N^{\beta}$, with $\alpha$ and $\beta$ the fit parameters and obtain $\alpha=0.47(2)$ and $\beta= -0.49(3)$. According to Eq.~(\ref{eq:tau}), $\tau\propto 1/U\eta$, leading to $U\eta\propto \sqrt{N}$. The dependence of the viscosity $\eta$ with the atom number depends on the variation of $U$ with $N$. 

In conclusion, we observe that the oscillating tunneling dynamics in a 1D bosonic Josephson junction relaxes to a phase-locked steady state. The timescale of the phenomenon remains unaffected by the amplitude and frequency of the oscillations. It is unchanged for a tunnel coupling ${J/h}$ between \SI{2}{Hz} and \SI{32}{Hz} and presents a dependence in approximately ${1/\sqrt{N}}$ with the atom number. The observed phase-locking is much faster than predicted in \cite{Torre2013,Foini2015,Foini2017} and leads to an equilibrium state with small fluctuations in phase and atom number imbalance. This questions the suitability of the quantum Sine-Gordon model to describe the out-of-equilibrium tunneling dynamics of two 1D superfluids confined in a harmonic potential. Ongoing work aims at determining the microscopic origin of the relaxation. This requires a deeper understanding of the relaxed state and an estimation of its energy. While this could in principle be deduced from ${J\langle\cos\phi\rangle}$ and from the final phase fluctuations, the latter are dominated by our imaging resolution and imaging shot noise in the current experiment. 

We are grateful to I. Mazets, T. Schweigler, G. Zaránd I. Lovas and E. Dalla Torre for helpful discussions and theoretical support. This research was supported by the ERC advanced grant QuantumRelax and by the Austrian Science Fund (FWF) through the project SFB FoQuS (SFB F40). M.P. and T.B. acknowledge the support of the Vienna Doctoral Program CoQuS. M.B. was supported by MC fellowship ETAB under grant agreement no.656530. E.D. acknowledges the support of the Harvard-MIT CUA, NSF Grant No. DMR-1308435 and of the AFOSR Quantum Simulation MURI, AFOSR Grant No. FA9550-16-1-0323.



\bibliography{biblio}

\clearpage
\onecolumngrid

\renewcommand{\thefigure}{S\arabic{figure}}
\setcounter{figure}{0}

\begin{center}
  \LARGE
  \textbf{Supplementary Materials} 
\end{center}



\section{Experimental preparation of the tunneling dynamics} 

\textbf{Generation of a 1D-BEC.}

An atom chip is used to trap $^{87}$Rb atoms in the $F=1, m_F=-1$ state of a magnetic trap elongated along the $z$-axis. The trap is characterized by the transverse frequencies ${\omega_{x,y}=2\pi\times}$\SI{3}{kHz} and the longitudinal frequency ${\omega_z=2\pi\times}$\SI{22}{Hz}. In this trap, the atoms are cooled down to degeneracy by radio-frequency evaporative cooling. The typical atom number at this stage is $4500$\,atoms, associated to a chemical potential $\mu/h=$\SI{5}{kHz}. The atom number was decreased in a set of data to investigate the effect of the atom number on the relaxation. In order to vary the atom number with limited change of temperature, we maintain a radio-frequency field of \SI{14}{kHz} above the trap bottom for a duration varying between \SI{0}{ms} and \SI{60}{ms} and evaporate the atoms heated by collisions. With this method, we vary the atom number by a factor of $6$ with negligible change of the temperature. \\

\textbf{Splitting.} 

The atom chip enables the generation of radio-frequency fields of tunable amplitude and orientation. We use this feature to perform a radio-frequency dressing of our trap and obtain an elongated double-well potential.  The symmetry of the trap, controlled by the field orientation, presents negligible fluctuations and a drift below 1$^\circ$ at the timescale required to obtain a data set. The barrier height is tuned by the field amplitude. The maximal current is ${I_{\text{max}}=\SI{80.25}{mA-pp}}$. To perform the splitting of the wave-functions, the dressing amplitude is ${0.65\times I_{\text{max}}}$, corresponding to a trap of frequencies ${\omega_x=2\pi\times\SI{1.7}{kHz}}$, ${\omega_y=2\pi\times\SI{1.7}{kHz}}$ and ${\omega_z=2\pi\times\SI{13}{Hz}}$. We perform a linear splitting ramp in \SI{21.5}{ms}. This duration gives the best number-squeezing factor ($\xi_N=0.57(6)$), relevant to slow down phase diffusion at best during the phase preparation. The typical phase diffusion rate we measured for $N\approx 2500$\,atoms is \SI{0.05(2)}{rad/ms} such that the phase coherence remains high after our state preparation. Another effect resulting from the splitting in a breathing mode excited by the change of trap geometry. When raising the tunnel barrier by RF-dressing, the longitudinal trap frequency $\omega_z$ first decreases, resulting in a trap decompression, before increasing again. This, associated to the splitting in two components of the original condensate, results in a slow breathing motion at a typical frequency of \SI{30}{Hz}. The timescale of the breathing is slow compared to the decay time of the oscillations but clearly appears on the fluorescence pictures as an increase of the length of the interference patterns along the $z-$axis (Fig.~3).
\\

\textbf{Phase imprinting.}

In the trap of negligible coupling, we imprint a global phase difference between the two wave-functions, named $\phi_0=\phi_L(0)-\phi_R(0)$.  For this, we realize a vertical displacement of one of the sites with respect to the other to induce an energy difference. 
This stage takes 3~ms, during which the the energy difference is ramped up from $\approx0$ to a final value $\epsilon$ and down again to $\approx0$ to recover a symmetric trap. We prepare initial phases between $-0.2\pi$ and $-0.8\pi$.\\

\textbf{Recoupling to the tunneling trap.} 

After the phase preparation, the barrier height is linearly decreased in \SI{3}{ms} toward the trap of study to perform a recoupling between the wave-functions.  In the scope of this study, the dressing amplitude of the final trap was varied between $0.525\times I_\text{max}$ and $0.6\times I_\text{max}$. The corresponding transverse frequency varies from $\omega_x=2\pi\times\SI{1.2}{kHz}$ to $\omega_x=2\pi\times\SI{1.5}{kHz}$ and $\omega_y$ varies from $\omega_y=2\pi\times\SI{1.6}{kHz}$ to $\omega_y=2\pi\times\SI{1.7}{kHz}$. This corresponds to a well spacing between $w=\SI{1.18}{\micro m}$ and $w=\SI{1.78}{\micro m}$. The longitudinal frequency varies between $\omega_z=2\pi\times\SI{12}{Hz}$ and $\omega_z=2\pi\times\SI{13}{Hz}$. During the recoupling, the tunnel coupling value $J$ is not instantaneously turned from $\approx0$ to its final value, but ramps up. Hence, the system undergoes a tunneling dynamics in a continuously changing trap. This is one reason why the data at time $t=\SI{0}{ms}$ presented in Fig. 2 do not correspond to the prepared state: the imbalance is not zero and the initial phase is not the maximal amplitude of the phase oscillation. In addition, the measurement procedure constrains us to lose approximately \SI{0.75}{ms} of the dynamics, due to a time offset detailed in the following part.\\

\section{Correction of technical offsets}
The oscillations in phase and imbalance presented in the paper are corrected from three technical offsets that we take into account as additional fit parameters. This section presents the nature and origin of these offsets as well as their typical values.\\

\textbf{Temporal offset.} 

A time shift between the oscillations in phase and imbalance is introduced by the measurement protocol, which differs for the phase and the imbalance. 
In the case of the relative phase measurement, we abruptly switch off the double-well potential and realize a time-of-flight imaging. The clouds fall and expand under the effect of inter-atomic interactions, primarily along the transverse direction. After \SI{46}{ms} of free expansion, the wave-packets fully overlap and interference fringes are visible. We then measure the density profiles using a light-sheet fluorescence detection system. The imbalance measurement relies on the same principle but requires an additional separation stage to prevent the overlap of the wave-packets. Before switching off the trap, the barrier height is raised to its highest value in \SI{1}{ms}, which is much faster than the inverse Josephson frequency. It gives a momentum to the atoms in the direction opposite to the barrier and excites a dipolar oscillation. After \SI{0.2}{ms}, the atoms reach the maximum amplitude of the oscillation and we rapidly switch off the dressing amplitude ($\approx \SI{360}{\micro s}$). The atom clouds suddenly experience a single trap and acquire some additional velocity toward each other. After a quarter of one oscillation, this velocity is maximal and the trap is switched off to perform a time-of-flight measurement. In this process, the clouds cross each other, but the number of collisions is negligible.

The duration of the measurement sequence before switching off the trap is the same for both the phase and the imbalance. However, while the imbalance measurement involves a quasi-instantaneous decoupling by raising the tunnel barrier, the phase measurement keeps the atoms in the coupled trap. It results in a systematic time shift $dt$ between the phase oscillations and the imbalance oscillations. The time shift is evaluated using the fit model presented in the paper, which requires to fit the phase and imbalance simultaneously. An interpolation of the imbalance data is required as the time shift is not necessarily a multiple of the experimental time step. We evaluate the time shift for each data set and the value is displayed in the TABLE~\ref{Tab:parameters_values}. The imbalance data presented in the paper are measured. The fit is performed on the measured phase data and on the interpolated imbalance data shifted by $-dt$. The data in the phase portrait representation are the measured phase values and the interpolated imbalance values.

\begin{figure}[h] 
\includegraphics[width=0.8\columnwidth]{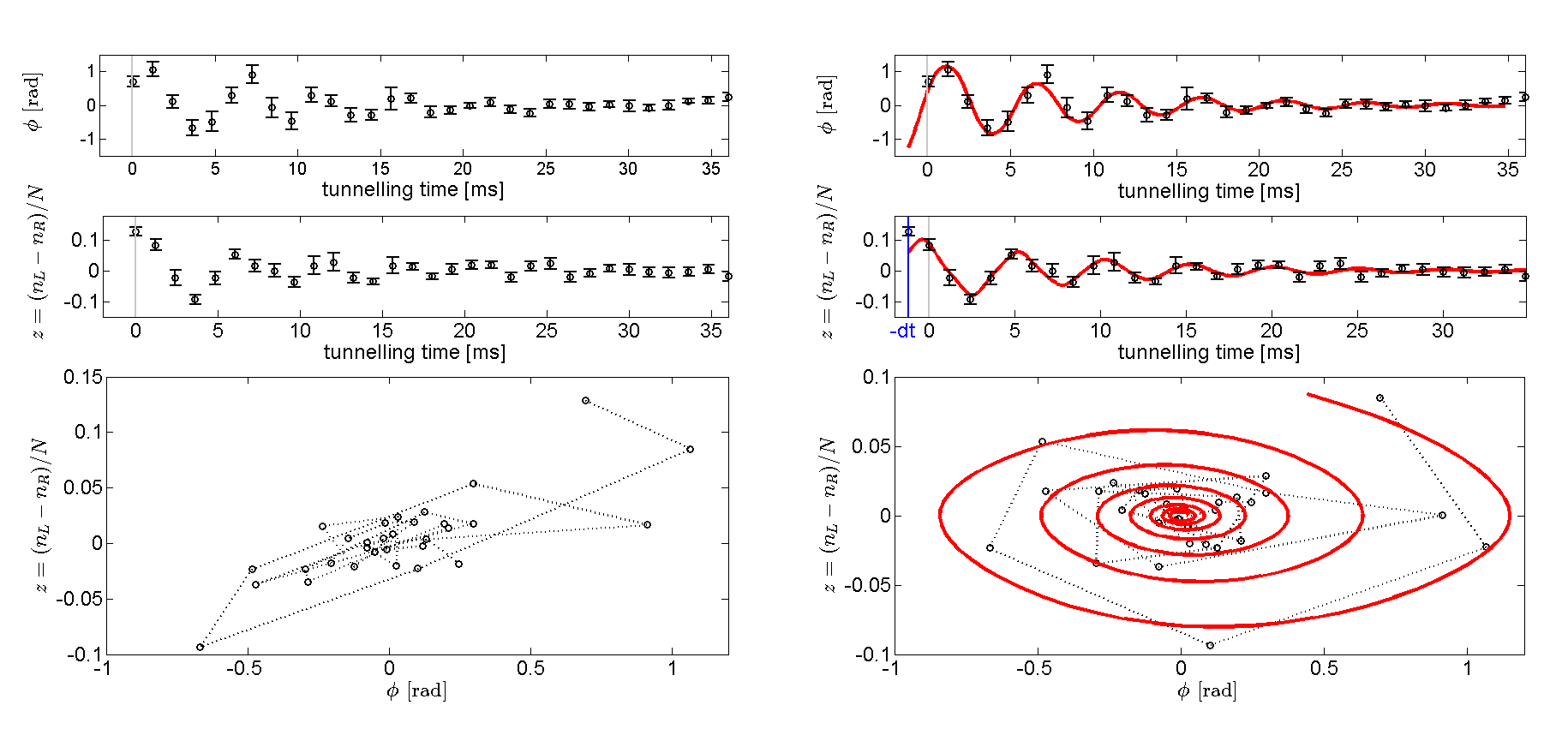}
\caption{\label{Fig:phase amplitude} Time shift correction.\textbf{ Left:} Time evolution of the phase and imbalance before time shift correction. The fit is not possible and the pairs ($n$,$\phi$) in the phase portrait present a tilt characteristic of a time shift. \textbf{ Right:} Time evolution of the phase and imbalance where the imbalance oscillation is shifted by $-dt$. The time shift in this case is $dt=\SI{0.75}{ms}$.}
\end{figure}

\textbf{Phase offset} 

The measured phase is extracted for each experimental realization from the interference profiles integrated along the longitudinal axis. We assume that the interference signal is a generic envelope $g(x)$ modulated by an interference term, such that the transverse density reads:

\begin{equation}
\rho(x)\approx g(x)[1+C\cos(k_0(x-x_0)+\phi)] \label{eq:transverse density}
\end{equation}
with $k_0$ the wave-vector, $C$ the fringe contrast, $\phi$ the phase and $x_0$ the reference pixel defining the phase origin. $x_0$ corresponds to the maximum of the generic envelope. An error on the determination of $x_0$ results in a systematic error on the phase analysis and a phase offset. $x_0$ is difficult to determine on an individual realization. To achieve the best determination of the envelope maximum, we assume that its position is fixed during the time required to measure a data set. It allows us to average the fluorescence pictures obtained for different tunneling times. As the phase oscillation corresponds to an oscillation of the fringes in the generic envelope, averaging fluorescence pictures of phases of large amplitude oscillations provides the best reconstruction of the envelope. Fitting the averaged profile with Eq.~(\ref{eq:transverse density}) assuming a Gaussian envelope gives the best evaluation of $x_0$. Once $x_0$ is determined, the most general approach to determine the phase consists in taking the Fourier transform of Eq.~(\ref{eq:transverse density}):
\begin{equation}
F(k)= \sqrt{2\pi} G(k) * \left[\delta(k)+\frac{C}{2}\left(e^{i\phi}\delta(k-k_0)+e^{-i\phi}\delta(k+k_0)\right) \right].
\end{equation}
where $G(k)$ is the Fourier transform of the generic envelope, $\delta(k)$ the Dirac distribution and $*$ the convolution product.  The modulus of $G(k)$ consists in a middle peak corresponding to the envelope and two side bands centered on $k_0$ and $-k_0$. The phase is computed by averaging the complex argument of $G(k)$ over one side band:
 \begin{equation}
\phi_{FT}=\arg \left[ \int_{k_0-\Delta k/2}^{k_0+\Delta k/2}G(k)\right].
 \end{equation}

Despite the careful evaluation of $x_0$, a small residual offset on the phase oscillations can still be present, for instance because of the finite pixel size, which limits our precision. This offset is extracted from the fit parameter $\bar{\phi}$
\begin{equation}
\phi(t)= 2\arcsin \left[ \sin \left( \frac{\Phi_0}{2}\right)e^{-t/\tau}\times\JacobiSN{\omega t+\phi'}{\sin\left(\frac{\Phi_0}{2}\right) e^{-t/\tau} } \right]+\bar{\phi}. \label{eq:phi}
\end{equation}
The value of the phase offset is of the order of \SI{0.2}{rad} which typically corresponds to a fringe displacement of one pixel. \\

\textbf{Imbalance offset} 

A small asymmetry between the two sites of the double-well can be present as we do not calibrate the trap symmetry before each measurement. This goes beyond the pendulum analogy and must be corrected to present the oscillations in the phase portrait representation as in Fig.~2. We can show that an energy difference $\epsilon$ between the two sites of the double-well results in an offset of the atomic imbalance $\bar{n}$ defined by: 
\begin{equation}
\bar{n}=-\frac{\epsilon}{\hbar}\frac{1}{\omega}\frac{N_0}{2\sin\left(\frac{\Phi_0}{2}\right)},
\end{equation}
with $\omega$ the plasma frequency and $N_0$,$\Phi_0$ the amplitudes of the oscillations in imbalance and phase. We take it onto account as an additional fit parameter $\bar{n}$ such that the fit function of the imbalance reads: 
\begin{equation}
n(t)=\frac{N_0}{\omega\times 2\sin(\frac{\Phi_0}{2})}\dot{\phi}(t)+\bar{n}.
\end{equation}

Typically, $\bar{n}$ is below 0.03. When comparing the damping in a symmetric and slightly asymmetric trap ($\bar{n}\approx 0.1$), we observe no significant effect of the asymmetry on the damping time. \\

\section{Fit results}
For each set of data, we display in TABLE~\ref{Tab:parameters_values} the values of the atom number deduced from the fluorescence pictures, the values of the technical offsets, the fit parameters extracted from Eqs.~(8)(9) and the corresponding parameters of the 2-mode model deduced from Eqs.~(10)(11)(12). The amplitude of the imbalance oscillations $N_0$ is hard to determine as it is very small and depends strongly on $dt$. As $N_0$ impacts strongly $U/h$, we constrained $N_0$ (within the error bars of $n$) to maintain a value of U consistent between the various data sets. The impact of this constrains on the fit residuals is negligible. The constrained values are indicated in bold.

\begin{table}[h]
\centering
       \begin{tabular}{|l|c||c|c||c|c|c|c||c|c|c|c||c|}
       \hline
			\multirow{2}{*}{Scan ref} & Atom &  \multicolumn{2}{c||}{Technical offsets} &  \multicolumn{4}{c||}{Fit parameters}&  \multicolumn{4}{c||}{2-mode model}& \multirow{2}{*}{Figures}\\
			
				\cline{3-12} 
				
		 & number &  dt\,[ms] & $\bar{n}$&  $\Phi_0$\,[rad]& $N_0$& $\tau$\,[ms]& $\omega$\,[Hz]&   $U/h$\,[Hz]& $J/h$\,[Hz]& $\eta$& $\epsilon/h$\,[Hz]&  \\			

		 \hline
		 \hline
	150216$\_$22& 3400 & 0.5 & 0.032(5)& 1.14(2)& \textbf{0.16(2)}& 9.0(2)& 2697(5)& 0.86(9)& 32(3)& 41(5)& 93(18)& Fig.\,4(b) \\ \hline
    150216$\_$6667& 3200 & 0.9 & -0.016(7)& 1.24(2)& \textbf{0.13(2)}& 9.9(3)& 2394(5)& 0.86(13)& 21(3)& 37(6)& 54(24)& Fig.\,4(b)\\ \hline
	150216$\_$24& 3400 & 1.0 & -0.025(7)& 1.59(3)& 0.12(3)& 7.9(2)& 1626(6)& 0.94(19)& 11(2)& 43(8)& 79(27)& Fig.\,4(b)\\ \hline
	150216$\_$6165& 3200 & 1.1 & -0.009(5)& 1.32(3)& 0.11(2)& 9.8(2)& 1248(4) & 0.71(15)& 8(2)& 46(10)& 14(13)& Fig.\,4(b),(c)\\ \hline
	150216$\_$23& 3300 & 1.8 & -0.007(3)& 1.53(7)&\textbf{ 0.07(3)}& 7.6(4)& 607(6)& 0.68(32)& 2(1)& 62(30)& 16(19)& Fig.\,4(b)\\ \hline
	150306$\_$380& 4500 & 1.3 & 0.008(3)& 1.89(7)& \textbf{0.12(4)}& 7.9(4)& 1337(8)& 0.67(25)& 8(3)& 61(23)& -24(24)& Fig.\,4(c)\\ \hline
	150304$\_$170& 2500 & 1.5 & 0.036(4)& 1.49(4)& 0.12(2)& 9.0(2)& 965(3)& 0.71(11)& 7(1)& 50(8)& -65(12)& Fig.\,4(c)\\ \hline
	150304$\_$1730 & 1700& 1.5 & 0.048(8)& 1.58(5)& 0.14(2)& 11.8(5)& 761(3) & 0.72(12)& 6(1)& 37(7)& -58(15)& Fig.\,4(c)\\ \hline
	150304$\_$1760 & 1250& 1.1 & 0.055(6)& 1.26(6)& 0.14(2)& 15.1(6)& 701(3) & 0.76(11)& 7(1)& 28(4)& -52(10)& Fig.\,4(c)\\ \hline
	150307$\_$4701& 750 & 1.5 & -0.003(8)& 0.66(4)& 0.09(2)& 15.5(1.5)& 548(6) & 0.80(20)& 6(2)& 26(7)& 2(4)& Fig.\,4(a),(c)\\ \hline
	150307$\_$4700& 750 & 1.7 & 0.001(9)& 1.30(5)& \textbf{0.17(3)}& 13.7(7)& 554(4)& 0.86(13)& 6(1)& 27(4)& -1(5)& Fig.\,4(a)\\ \hline
	150307$\_$4710& 750 & 1.6 & -0.002(23)& 1.77(16)& \textbf{0.19(6)}& 15.4(1.6)& 502(7)& 0.88(30)& 5(2)& 23(8)& 1(15)& Fig.\,4(a)\\ \hline

				\end{tabular}

\caption{Values of the experimental parameters, technical offets, fit parameters and deduced parameters of the 2-mode model for the data sets presented in Fig.~(4).\label{Tab:parameters_values}}
\end{table}

\textbf{Uncertainties and error bars:} 

The mean values of the phase and imbalance are typically computed using 6 repeats per tunneling time.  
The error bars of Fig.~2 are computed as standard error of the mean i.e. as standard deviation assuming a Gaussian distribution divided by the square root of the number of realizations. The fits are realized on the mean values without accounting for the error bars. A weighted fit gives comparable results but made the determination of $N_0$ difficult. The uncertainties on the fit parameters corresponds to the 95$\%$ confidence interval. The uncertainties on $U/h, J/h, \eta$ and $\epsilon/h$ are deduced from error propagation method.\\



\end{document}